\documentclass[prd,aps,twocolumn,nofootinbib,preprintnumbers,superscriptaddress,balancelastpage,longbibliography]{revtex4-2}
\usepackage[utf8]{inputenc}
\usepackage{graphicx}
\graphicspath{ {./figures/} }
\usepackage{hyperref}
\usepackage{slashed}
\usepackage{aas_macros}
\usepackage{float}
\usepackage{subfigure}
\usepackage{multirow}
\usepackage{amsmath}
\usepackage{array} % for defining a new column type
\usepackage{varwidth} %for the varwidth minipage environment
\usepackage{tabularx}
\usepackage{braket}
\usepackage[dvipsnames]{xcolor}
\usepackage{listings}
\usepackage[normalem]{ulem}

\hypersetup{
     colorlinks   = true,
     citecolor    = SeaGreen,
     urlcolor     = SeaGreen,
     linkcolor    = SeaGreen
}

\begin{document}

\title{Every Nearby Energetic Pulsar Is Surrounded by a Region of Inhibited Diffusion}

\author{Isabelle John}
\thanks{{\scriptsize Email}: \href{mailto:isabelle.john@unito.it}{isabelle.john@unito.it}; ORCID: \href{https://orcid.org/0000-0003-2550-7038}{0000-0003-2550-7038}}
\affiliation{Dipartimento di Fisica, Universit\`a degli Studi di Torino, via P.\ Giuria, 1 10125 Torino, Italy}
\affiliation{INFN -- Istituto Nazionale di Fisica Nucleare, Sezione di Torino, via P.\ Giuria 1, 10125 Torino, Italy}

\author{Tim Linden}
\thanks{{\scriptsize Email}: \href{mailto:linden@fysik.su.se}{linden@fysik.su.se};  ORCID: \href{http://orcid.org/0000-0001-9888-0971}{0000-0001-9888-0971}}
\affiliation{Stockholm University and The Oskar Klein Centre for Cosmoparticle Physics, Alba Nova, 10691 Stockholm, Sweden}
\affiliation{Erlangen Centre for Astroparticle Physics (ECAP), Friedrich-Alexander-Universität \\ Erlangen-Nürnberg, Nikolaus-Fiebiger-Str. 2,
91058 Erlangen, Germany}

\begin{abstract}
\noindent The H.E.S.S. telescope has recently detected the total electron-plus-positron ($e^+e^-$) flux up to 40~TeV, finding it to be a featureless and steeply-falling power-law above 1~TeV. This result is in stark tension with standard one-zone models of pulsar $e^+e^-$ injection and diffusion, which predict a hard-spectrum signal above $\sim$10~TeV. We model the local pulsar population, and find 21 sources that would each \emph{individually} overproduce the H.E.S.S. $e^+e^-$ flux in a one-zone diffusion model. We conclude that \emph{every} energetic pulsar younger than $\sim$500~kyr must be surrounded by a region of inhibited diffusion (\emph{e.g.,} a supernova remnant, pulsar wind nebula, or TeV halo) that prevents the transport of these $e^+e^-$ to Earth. Because the high-electron density in these regions produces bright synchrotron and inverse-Compton emission, we conclude that all nearby pulsars are detectable as (potentially unassociated) radio, x-ray or $\gamma$-ray sources. 
\end{abstract}

\maketitle

\section{Introduction}
Observations by the High-Energy Stereoscopic System (H.E.S.S.) have recently extended our detection of the combined electron-plus-positron (hereafter, $e^+e^-$) flux up to an energy of 40~TeV. The $e^+e^-$ flux has a sharp break at an energy of $\sim$1~TeV, above which the emission is characterized by a smooth and rapidly falling power-law, with $dN/dE \propto E^{-4.5}$~\cite{HESS:2024etj}.

This observation strongly constrains the production and propagation of $e^+e^-$ pairs from pulsars, which have long been considered to provide the dominant contribution to the local positron flux measured by PAMELA~\cite{PAMELA:2013vxg} and AMS-02~\cite{PhysRevLett.122.041102, PhysRevLett.122.101101} between $\sim$10--300~GeV~\cite{Hooper:2008kg, Profumo:2008ms, Malyshev:2009tw, Linden:2013mqa, Hooper:2017gtd, Orusa:2021tts, Cholis:2021kqk, Orusa:2024ewq}. As shown in Figure~\ref{fig: diffusion models}, the $e^+e^-$ spectrum from nearby pulsars in a standard one-zone model has two components. The first is the steep rise at low energies that corresponds to electrons with cooling timescales ($t_c$) longer than the pulsar age ($t_\text{PSR})$. This component has a cutoff produced by the cooling of very high-energy electrons to a critical energy where $t_c$ = $t_\text{PSR}$. The second, high-energy, component stems from recently produced $e^+e^-$ that propagate to Earth before cooling. The relative energy-flux of each feature depends on the pulsar age, spindown time, injection spectrum, and distance, but are roughly similar.

By normalizing the GeV flux from pulsars to AMS-02 data, we can predict the H.E.S.S. TeV $e^+e^-$ flux. The soft spectrum of the H.E.S.S. data then strongly constrains some convolution of the TeV $e^+e^-$ injection rate from pulsars or the propagation of these pairs to Earth. Such a constraint has previously been applied to argue that $e^+e^-$ propagation must be inhibited near bright pulsars such as Vela and Geminga~\cite{Profumo:2018fmz, Huang:2018phb, Tang:2018wyr, Evoli:2018aza, Lopez-Coto:2018ksn, DiMauro:2019yvh, Fang:2019ayz, Manconi:2020ipm, Martin:2022hrx, Schroer:2023aoh, Fang:2023xla}. 

In this \emph{article} we significantly extend this result to prove that diffusion must be inhibited around \emph{every} pulsar (seen or unseen) that is either powerful enough, or close enough, to contribute to the high-energy $e^+e^-$ flux at Earth. Moreover, our analysis can be applied to any ensemble of pulsars that combine to contribute to the rising AMS-02 positron flux, regardless of their individual contributions. The argument proceeds as follows:

\begin{figure}[tbp]
    \centering
    \includegraphics[width=\linewidth]{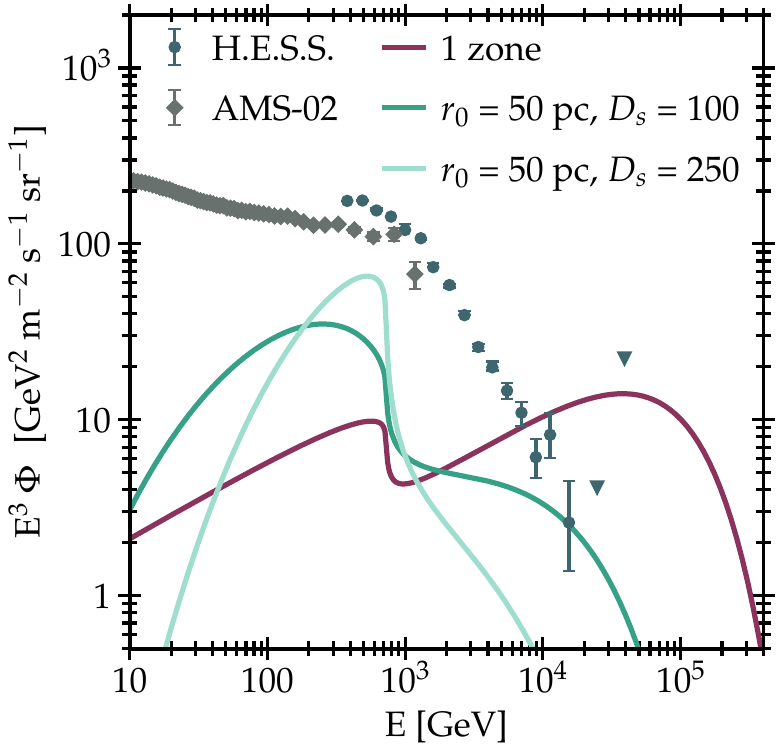}
    \vspace{-0.5cm}
    \caption{The $e^+e^-$ flux from a middle-aged pulsar (500~kyr old, 150~pc away) compared to H.E.S.S. (dark blue)~\cite{HESS:2024etj} and AMS-02 (gray)~\cite{PhysRevLett.122.041102, PhysRevLett.122.101101} data. The violet line corresponds to a one-zone diffusion model that overproduces the H.E.S.S. flux. Medium and light green represent two-zone diffusion models, where diffusion is inhibited around the pulsar. This suppresses the high-energy flux and agrees with data.}
    \label{fig: diffusion models}
\end{figure}

\begin{enumerate}
    \item [(1)] Recent observations of TeV halos~\cite{Linden:2017vvb, HAWC:2017kbo, HESS:2023sbf, LHAASO:2023rpg} provide very strong evidence that pulsars produce the positron excess~\cite{Hooper:2017gtd, Profumo:2018fmz, Fang:2018qco}, establishing them as the predominant $e^+e^-$ source near 1~TeV.
    
    \item [(2)] Young pulsars produce most of their $e^+e^-$ while rapidly spinning down soon after they are born~\cite{Profumo:2008ms}. However, the observation of bright TeV halos around Geminga and Monogem with $\gamma$-rays exceeding 50~TeV~\cite{2024ApJ...974..246A} establishes two key facts: (1) there is no a sharp cutoff in the pulsar $e^+e^-$ injection at a few TeV, as $e^+e^-$ up to $\sim$100~TeV are needed to produce the $\gamma$-ray spectra, and (2) this $e^+e^-$ acceleration must continue to late times, as these $e^+e^-$ pairs cool in $\sim$10~kyr. Recent HAWC, LHAASO, and H.E.S.S. measurements of TeV halos confirm that this phenomenon is generic, and not isolated to Geminga and Monogem~\cite{Sudoh:2019lav, LHAASO:2021crt, HAWC:2021dtl, HESS:2023qlw}.
        
    \item [(3)] In one-zone diffusion models, $e^+e^-$ from sources within $\mathcal{O}$(1~kpc) of Earth can efficiently reach Earth before cooling, producing a bright $\sim$10~TeV $e^+e^-$ flux with a total power that is a few percent the peak pulsar power near the cutoff produced by $e^+e^-$ cooling~\cite{Schroer:2023aoh}. For many nearby pulsars, this $e^+e^-$ flux will overproduce the H.E.S.S. data.

    \item[(4)] One method for decreasing the $e^+e^-$ flux at Earth would be to produce a bubble of \emph{locally} inhibited diffusion surrounding the solar position~\cite{HAWC:2017kbo}. However, this provides a poor fit to the data for several reasons: (1) strongly inhibited diffusion near the Sun will prevent all $e^+e^-$ from reaching the Earth, at odds with H.E.S.S. data~\cite{Hooper:2017tkg}, (2) a change in local diffusion will produce a common spectral cutoff that equally applies to all sources, producing a common spectral bump and fall-off (see Fig.~\ref{fig: diffusion models}) that does not fit the smooth power-law observed by H.E.S.S., (3) there are no known sources that could produce locally inhibited diffusion, while TeV halo observations indicate that many pulsars inhibit diffusion locally, (4) if locally inhibited regions of diffusion were common far from sources, it would significantly affect GeV secondary-to-primary ratios, significantly altering fits to AMS-02 data~\cite{Hooper:2017gtd}. 
\end{enumerate}

Based on these lines of evidence, there is only one remaining solution. Every pulsar that produces high-energy $e^+e^-$ pairs must \emph{also} produce a significant region of low diffusion, which prevents these $e^+e^-$ pairs from leaking into the interstellar medium (ISM) unperturbed. The origin of the inhibited diffusion is unclear, as both supernova remnants (SNR) and pulsar wind nebulae (PWNe) have long been assumed to inhibit diffusion in regions surrounding very young ($\lesssim 50$~kyr) pulsars~\cite{2012A&ARv..20...49V, Gaensler:2006ua, Mitchell:2022cpw}. However, by comparing current H.E.S.S. constraints against the ATNF pulsar catalog~\cite{Manchester:2004bp, ATNF_website}, we can confirm that these inhibited diffusion regions continue to at least $\sim$500~kyr, meaning that inhibited diffusion (most likely due to TeV halos) remains ubiquitous among middle-aged pulsars. Notably, these results are supported by $\gamma$-ray observations of middle-aged pulsars, which often show evidence of inhibited diffusion by factors of 100--1000 on scales of 25--50~pc~\cite{2024ApJ...974..246A}.

The ubiquity of inhibited diffusion around pulsars lets us make a second strong claim --- all sufficiently nearby and energetic pulsars (roughly within $\sim$0.5~kpc for a 300~kyr pulsar) have either already (or will soon be) detected. The inhibition of diffusion means that any such pulsar will produce bright extended TeV $\gamma$-ray emission independent of its beam orientation. The only caveats are: (1) very dim pulsars whose TeV $e^+e^-$ falls outside the normal pulsar population, (2) pulsars in regions that have not yet been covered by TeV instruments (primarily in the southern hemisphere), and (3) bright TeV sources that are known, but not currently associated to pulsars.

\section{Diffusion Model}\label{sec: diffusion model}
We compute the $e^+e^-$ flux from each pulsar using one-zone and two-zone analytic diffusion models~\cite{Hooper:2008kg, Hooper:2017gtd, Osipov:2020lty, Schroer:2023aoh}. Given a distance $r$ from the pulsar to Earth and a pulsar age $t_\text{PSR}$, the $e^+e^-$ flux is,

\begin{equation}\label{eq: pulsar flux}
\Phi\left(E, r, t_\text{PSR}\right) = \frac{c}{4\pi} \int_0^{t_\text{PSR}} \frac{E_0^2}{E^2} \; Q\left(E_0, t'\right) \mathcal{H}\left(r, E_0, t'\right) \, dt',
\end{equation}

\noindent where $Q\left(E_0, t'\right)$ is the $e^+e^-$ source term, $\mathcal{H}\left(r, E_0, t'\right)$ models e$^+$e$^-$ diffusion in each model, and $E$ and $E_0$ are the current and initial energy of an $e^\pm$ that was injected a time $t = t_\text{PSR} - t'$ ago, and are related via 

\begin{equation}\label{eq: initial energies}
    t = - \int_E^{E_0} \frac{1}{b(E')\, E'^2} \,dE'.
\end{equation}

\noindent The factor $b(E)$ denotes the $e^\pm$ energy losses, given by ${dE/dt = - b\left(E\right) E^2 }$, where $b(E)$ depends on the magnetic field and interstellar radiation field (ISRF) that control the synchrotron and inverse-Compton rates, including a precise treatment of Klein-Nishina effects, which are particularly important at TeV energies~\cite{Hooper:2017gtd, John:2022asa}. We follow the analytic method in Ref.~\cite{John:2022asa} using the ISRF model of Ref.~\cite{Porter:2008ve}. These models assume that energy losses are continuous and ignore their stochasticity~\cite{John:2022asa}. However, this does not affect our results, which probe energies well above the spectral feature produced by $e^\pm$ cooling. For a constant $b(E) = b$, Eq.~\ref{eq: initial energies} simplifies to ${ E_0 = E / (1-E\,b\,t) }$.

The $e^+e^-$ source term, $Q(E_0, t)$, is given by

\begin{equation}\label{eq: source term}
Q\left(E_0, t\right) = \eta L\left(t\right) Q_0\left(E_0\right).
\end{equation}

\noindent where $\eta$ is the pulsar's efficiency of converting its spindown luminosity $L(t)$ into $e^+e^-$ pairs, and $Q_0$ the injection term. The source term, Eq.~\ref{eq: source term}, is normalized by the current pulsar luminosity, $L(t)$, through

\begin{equation}\label{eq: luminosity}
L\left(t\right) = L_0 \left(1 + \frac{t}{\tau}\right)^{-\frac{n+1}{n-1}},
\end{equation}

\noindent where $L_0$ is the initial pulsar luminosity, $\tau$ the spindown time scale, and $n = 3$ is the braking index for a dipole.

The $e^+e^-$ injection spectrum $Q_0\left(E_0\right)$ is

\begin{equation}\label{eq: injection spectrum}
Q_0\left(E_0\right) = E_0^{-\alpha} \exp{\left(-\frac{E_0}{E_\text{cutoff}}\right)},
\end{equation}

\noindent where $\alpha$ is the spectral index and $E_\text{cutoff}$ is the $e^\pm$ cutoff energy of the pulsar. The integral over time in Eq.~\ref{eq: pulsar flux} accounts for the continuous injection of $e^\pm$ by the pulsar.

In the \textit{one-zone diffusion} model~\cite{Hooper:2008kg, Hooper:2017gtd}, the diffusion factor $\mathcal{H}$ becomes

\vspace{-0.15cm}
\begin{equation}\label{eq: one-zone diffusion}
\mathcal{H} = \frac{1}{\left(4\pi\,\lambda^2\left(E_0, E\right)\right)^{3/2}} \exp{\left(-\frac{r^2}{4\,\lambda^2\left(E_0, E\right)}\right)}.
\end{equation}

\noindent where the diffusion distance scale, $\lambda$, is given by

\begin{equation}\label{eq: diffusion distance}
\begin{aligned}
\lambda^2\left(E_0, E\right) &= \int_{E}^{E_0} \frac{D\left(E'\right)}{-dE'/dt} \, dE' \\ &= \frac{D_1}{b E^{\left(1-\delta\right)} \left(1-\delta\right)} \left(1 - \left(1 - bEt\right)^{\left(1-\delta\right)}\right),
\end{aligned}
\end{equation}

\noindent where ${ -dE/dt = b(E)E^2 }$. Note that Eq.~6 is only formally true when b is independent of energy, $b(E) = b$. In the calculations of this paper, we take into account the energy dependence of $b(E)$ and include Klein-Nishina effects. Thus, we note that, while not an exact equality, our method accounts for the energy dependence of $b(E)$ and is a sufficiently accurate approximation of the calculation. We set the diffusion coefficient $D\left(E\right)$ via an energy scaling of the diffusion constant $D_1$ as \mbox{$D\left(E\right) = D_1 E^\delta$}.

In the \textit{two-zone diffusion} model, diffusion is inhibited in a region $r_0$ around the pulsar. We set the diffusion constant in the inner zone to be $D_0$ and in the outer zone to be $D_1$ with $D_0 < D_1$. Then $\mathcal{H}$ becomes~\cite{Osipov:2020lty, Schroer:2023aoh}

\begin{equation}
\mathcal{H}\left(r, E, t\right) = \int_0^\infty \frac{\xi e^{-\psi}}{\pi^2\lambda_0^2 \;r \left(A^2\left(\psi\right) +  B^2\left(\psi\right)\right)} \times \mathcal{D}\left(\psi\right) \,d\psi,
\end{equation}

\noindent where $\mathcal{D}\left(\psi\right)$ for the one-zone diffusion model is

\begin{equation}
\mathcal{D}_\text{1-zone}\left(\psi\right) = \sin{\left(\sqrt{\psi}\frac{r}{\lambda_0}\right)}
\end{equation}

\noindent and has the analytic solution given in Eq.~\ref{eq: one-zone diffusion}. In the two-zone diffusion model, where $ r \geq r_0$, this becomes

\begin{equation}
\begin{aligned}
\mathcal{D}_\text{2-zone}\left(\psi\right) &= A\left(\psi\right)\sin{\left(\sqrt{\psi}\frac{r\xi}{\lambda_0}\right)} \\
&+ B\left(\psi\right)\cos{\left(\sqrt{\psi}\frac{r\xi}{\lambda_0}\right)}.
\end{aligned}
\end{equation}

The factors $A$ and $B$ are defined as

\begin{equation}
\begin{aligned}
A\left(\psi\right) &= \xi\cos{\left(\chi\right)}\cos{\left(\xi\chi\right)} + \sin{\left(\chi\right)}\sin{\left(\xi\chi\right)} \\ 
&+ \frac{1}{\chi} \left(\frac{1-\xi^2}{\xi} \sin{\left(\chi\right)} \cos{\left(\xi\chi\right)}\right)
\end{aligned}
\end{equation}

\noindent and

\begin{equation}
B\left(\psi\right) = \frac{\sin{\left(\chi\right)} - A\left(\psi\right) \sin{\left(\xi\chi\right)}}{\cos{\left(\xi\chi\right)}},
\end{equation}

\noindent where $\chi = \sqrt{\psi}\frac{r_0}{\lambda_0}$ and $\xi = \sqrt{\frac{D_0}{D_1}}$. Note that for one-zone diffusion, $A=1$, $B=0$ and $\xi=1$. The diffusion distance in each zone is expressed in terms of $\lambda_0$, which is related to $D_0$, \textit{i.e.} the suppressed diffusion constant, in Eq.~\ref{eq: diffusion distance}. 

For standard diffusion in the interstellar medium, we choose $D_\text{ISM} = 2 \times 10^{28}$~cm$^2$/s at an energy of 1~GeV with a diffusion index $\delta = 0.4$~\cite{Korsmeier:2021bkw, DiMauro:2023jgg}. For two-zone diffusion, we set the size of the inner diffusion zone to $r_0 = 50$~pc and test a suppression factor $D_s = D_1/D_0$. We consider two cases, where $D_s = 100$ and 250 with $D_1 = D_\text{ISM}$. The size of the diffusion zone and the suppression factor are degenerate (when energy losses are neglected) via ${ \ell = \sqrt{6Dt} }$, where $\ell$ is the average distance an $e^\pm$ travels in a time $t$ given a diffusion coefficient $D$ when energy losses are negligible. The time that the $e^+e^-$ spend around the pulsar determines the suppression of the high-energy flux.

\section{Pulsar Model}\label{sec: pulsar model}
We use the ATNF pulsar catalog (v. 2.6.0)~\cite{Manchester:2004bp, ATNF_website}, and select every pulsar with age and distance estimates closer than 1.5~kpc and younger than 1~Myr. We use a generic pulsar model to simulate pulsars in our analysis. While in reality, each pulsar is expected to have different parameters that govern $e^+e^-$ injection and diffusion (such as initial spindown luminosity, braking index, injection spectrum), our results are consistent across a wide variety of ATNF pulsars, and thus changes in single systems do not significantly affect our results. We loosely base our generic pulsar model on Geminga~\cite{Hooper:2017gtd}, which is the best-measured middle-aged pulsar. For the pulsar injection spectrum, we use Eq.~\ref{eq: injection spectrum} with $\alpha=2.0$ and $E_\text{cutoff} = 100$~TeV. We note that our choice of the slope of the injection spectrum is slightly degenerate with the strength of required inhibited diffusion -- a softer injection results in a smaller excess of the H.E.S.S. data at high energies. While the exact injection spectrum is unknown, and likely strongly varies between sources, we have selected a generic benchmark model with $\alpha = 2$, which seems to represent many of the TeV pulsars and halos observed by current instruments~\cite{Alfaro:2026tak, Bao:2024rrg, HESS:2023qlw, us_upcoming}. In Eq.~\ref{eq: luminosity}, we set the initial spindown luminosity to $L\left(t=0\right)= 10^{38}$~erg/s~\cite{Linden:2017vvb} and use a spindown timescale of $\tau = 10$~kyr~\cite{HAWC:2017kbo}, with an $e^+e^-$ efficiency of $\eta=0.1$. We evaluate our model pulsars on a grid of ages spanning 5~kyr--1~Myr in bins of 5~kyr and distances that span 100--1500~pc with bins of 50~pc. We compare these results to the ATNF population.

\begin{figure}[tbp]
\centering
\begin{minipage}[t]{0.47\textwidth}
\includegraphics[width=0.86\textwidth]{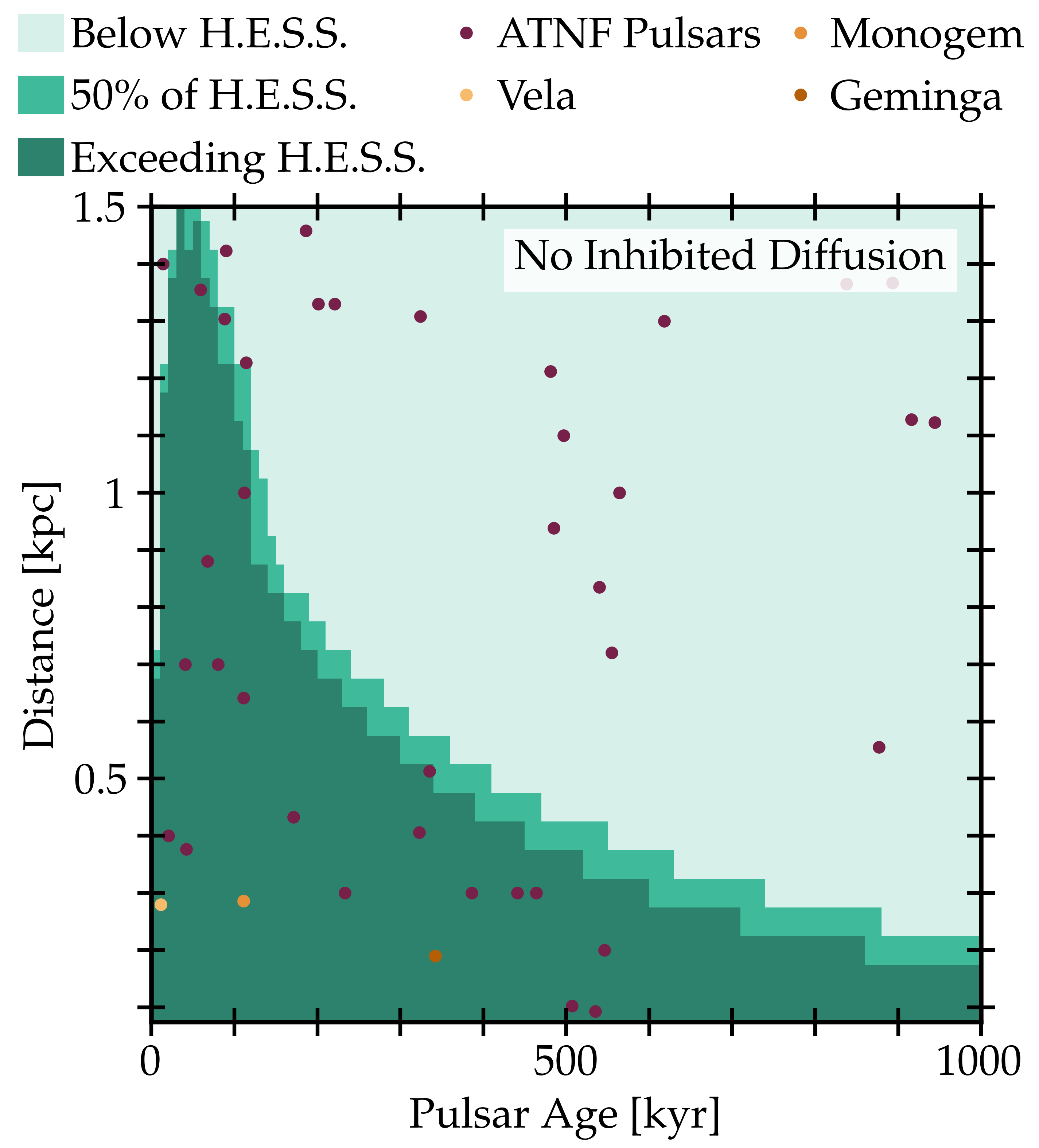} 
\end{minipage}
\hfill
\begin{minipage}[t]{0.47\textwidth}
\includegraphics[width=0.86\textwidth, trim={0 0 0 2.6cm},clip]{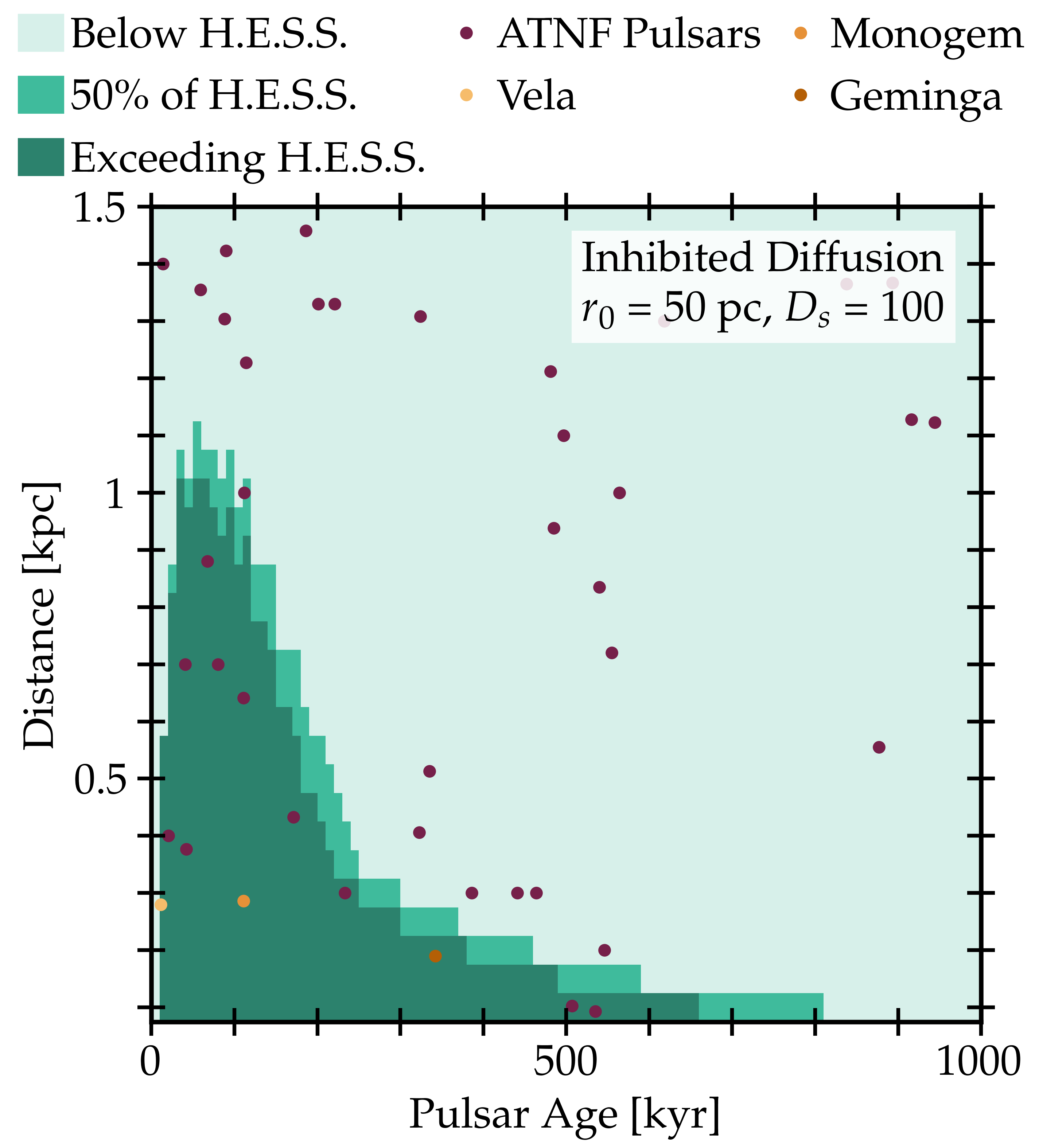} 
\end{minipage}
\hfill
\begin{minipage}[t]{0.47\textwidth}
\includegraphics[width=0.86\textwidth, trim={0 0 0 2.6cm},clip]{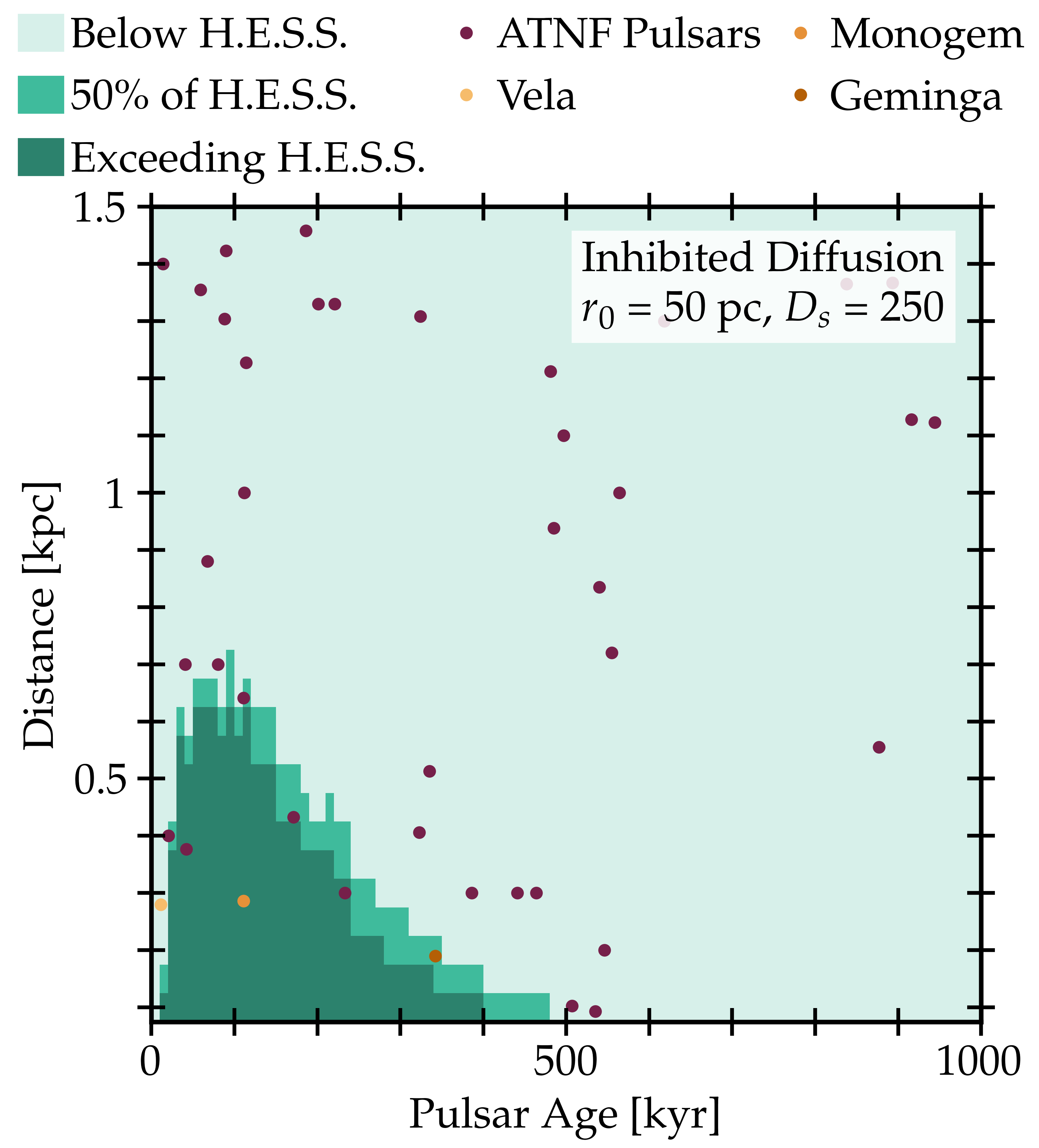} 
\end{minipage}
\vspace{-0.3cm}
\caption{The population of known pulsars that \emph{individually} overproduce the H.E.S.S. $e^+e^-$ data (dark green), account for 50\% of H.E.S.S. (medium green), and less (light green), for the one-zone diffusion model (top panel), two-zone model for \mbox{$r_0 = 50$~pc} with $D_s = 100$ (middle panel) and $D_s = 250$ (bottom panel). Dots correspond to known ATNF pulsars.}
\label{fig: pulsars vs HESS}
\end{figure}

\begin{figure}[tbp]
\vspace{-0.2cm}
\centering
\begin{minipage}[t]{0.47\textwidth}
\includegraphics[width=0.86\textwidth, trim={0 0 0 2.6cm},clip]{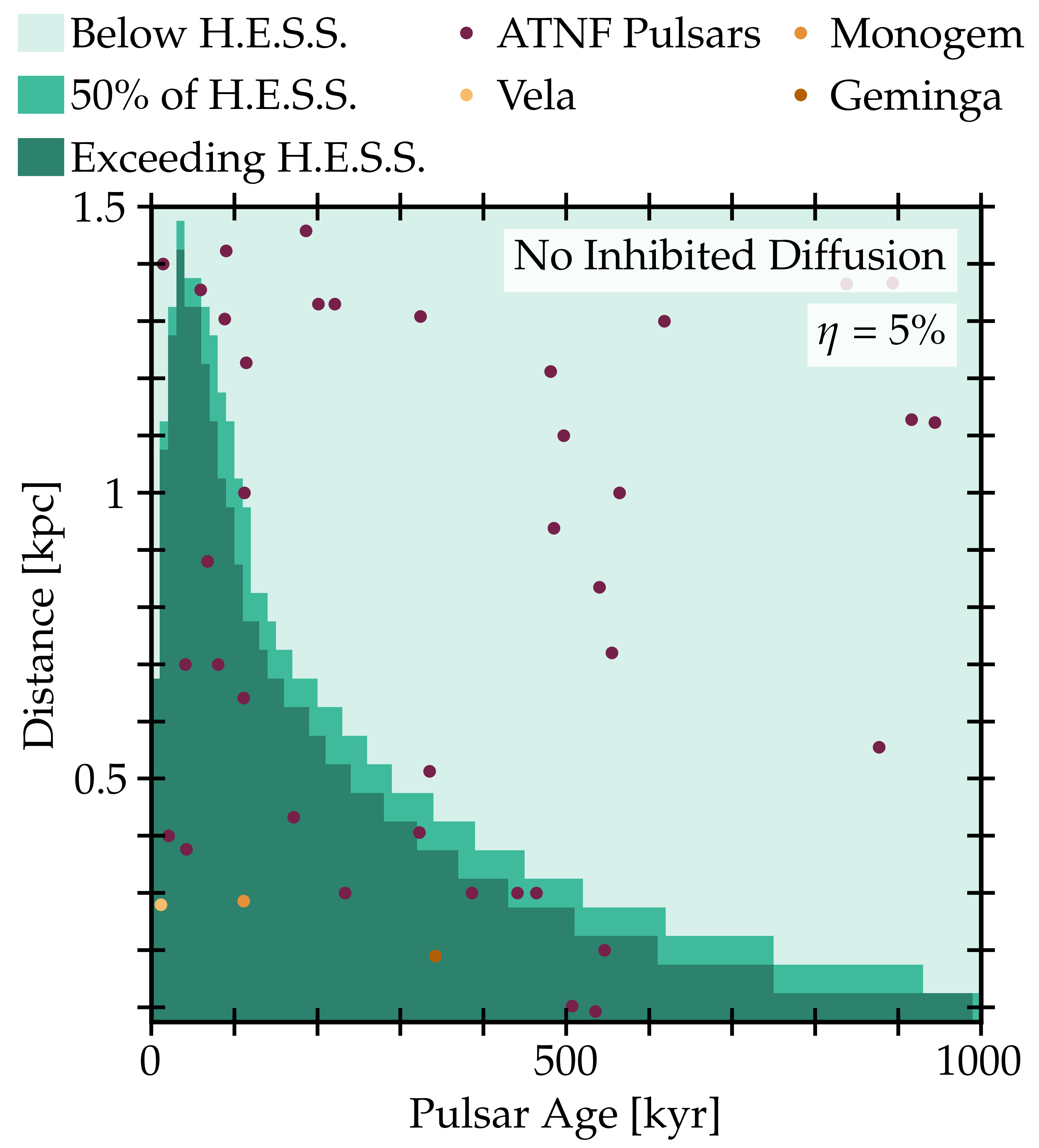} 
\end{minipage}
\vfill
\begin{minipage}[t]{0.47\textwidth}
\centering
\includegraphics[width=0.86\textwidth, trim={0 0 0 2.6cm},clip]{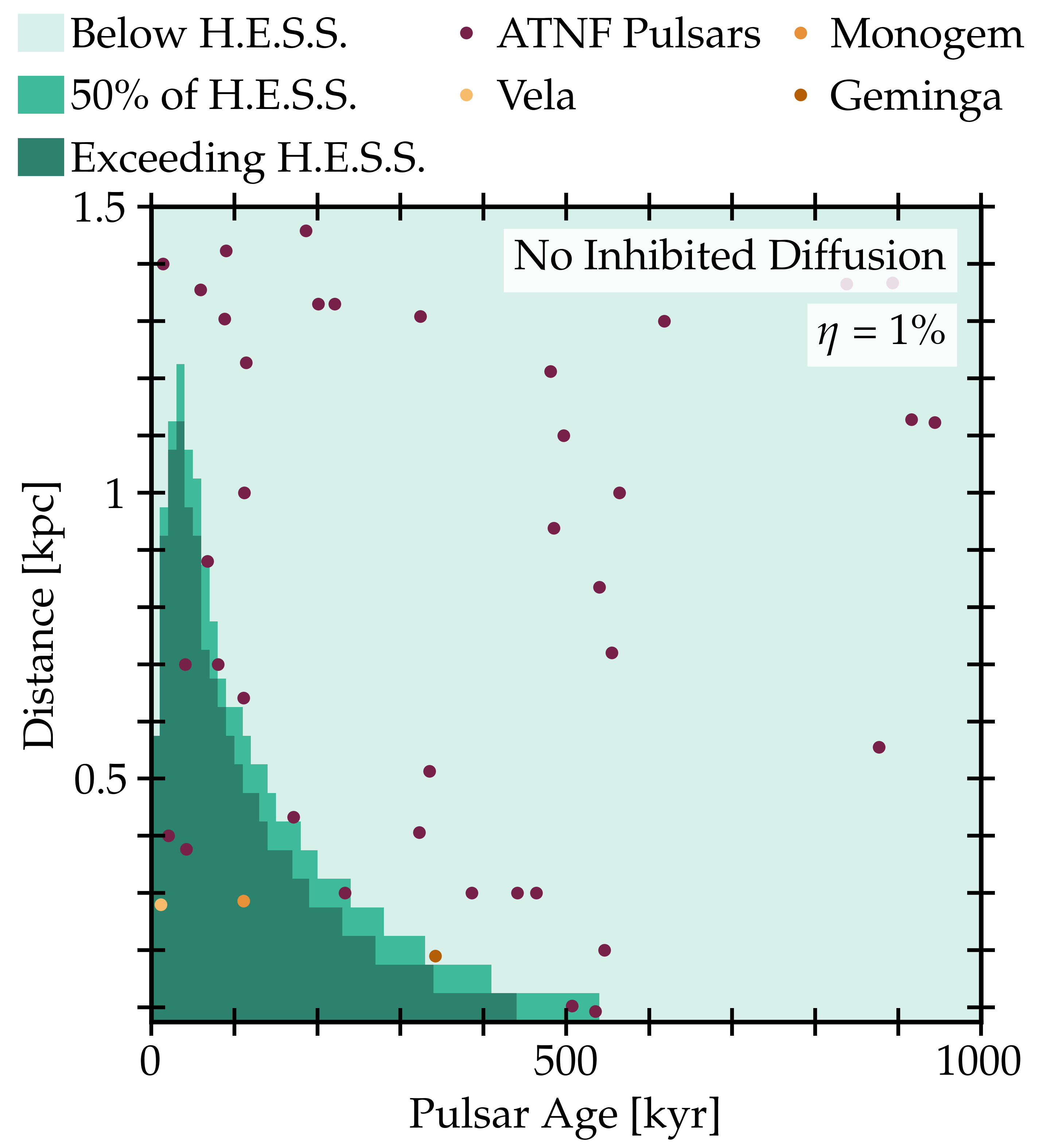} 
\end{minipage}
\vspace{-0.2cm}
\caption{Similar to Fig.~\ref{fig: pulsars vs HESS} but for pulsar efficiencies of $\eta = 5$\% (top) and $\eta = 1$\% (bottom) instead of 10\%.}
\label{fig: pulsars vs HESS 0.01 eff}
\end{figure}

\section{Results and Discussion}\label{sec: results and discussion}
In Fig.~\ref{fig: diffusion models}, we compare the H.E.S.S. $e^+e^-$ flux (dark blue)~\cite{HESS:2024etj} with our generic pulsar model at an age of 500~kyr and distance of 150~pc. The standard one-zone model (violet) has a high-energy flux that significantly exceeds the H.E.S.S. data. In comparison, the dark (light) green spectrum represents two-zone diffusion models where diffusion is suppressed by a factor of 100 (250) in a 50~pc region around the pulsar. These models have TeV $e^+e^-$ fluxes consistent with data.

Figure~\ref{fig: pulsars vs HESS} compares pulsars of different ages and distances against H.E.S.S. $e^+e^-$ data for three diffusion models: the one-zone model (top panel), the two-zone model with $r_0 = 50$~pc and $D_s = 100$ (middle) and $r_0 = 50$~pc and $D_s = 250$ (bottom). Violet dots represent ATNF pulsars~\cite{Manchester:2004bp, ATNF_website}, while Vela, Monogem and Geminga are shown in yellow, orange and dark orange, respectively. Pulsars in dark green regions have $e^+e^-$ fluxes that \emph{individually} overproduce H.E.S.S. data. Medium green regions represent pulsars that individually account for more than 50\% of the flux in a single H.E.S.S. energy bin. Light green regions indicate pulsars whose flux falls below H.E.S.S. data.  In the one-zone model, $\sim 21$ known pulsars overproduce the H.E.S.S. data. The number reduces to 12 in the two-zone model with a $D_s = 100$ suppression, while in the case of $D_s$~=~250 only three sources exceed the H.E.S.S. flux. We note that this model sets a lower limit on the suppression of diffusion, as many pulsar wind nebulae and TeV halos show evidence for diffusion to be inhibited by a factor of 1000 or more~\cite{DiMauro:2019hwn, 2024ApJ...974..246A}.

Figure~\ref{fig: pulsars vs HESS 0.01 eff} shows the same result as Fig.~\ref{fig: pulsars vs HESS} for models with pulsar efficiencies of $\eta = 5\%$ and 1\%. This reduces the flux from each pulsar and reduces the number that exceed H.E.S.S. data. However, we find that for a 5\% (1\%) efficiency, 15 (5) pulsars still exceed H.E.S.S data. Even for pulsars with much lower efficiencies, inhibited diffusion is needed to be consistent with H.E.S.S.

\begin{figure}[tbp]
    \centering
    \vspace{-0.3cm}
    \includegraphics[width=0.95\linewidth]{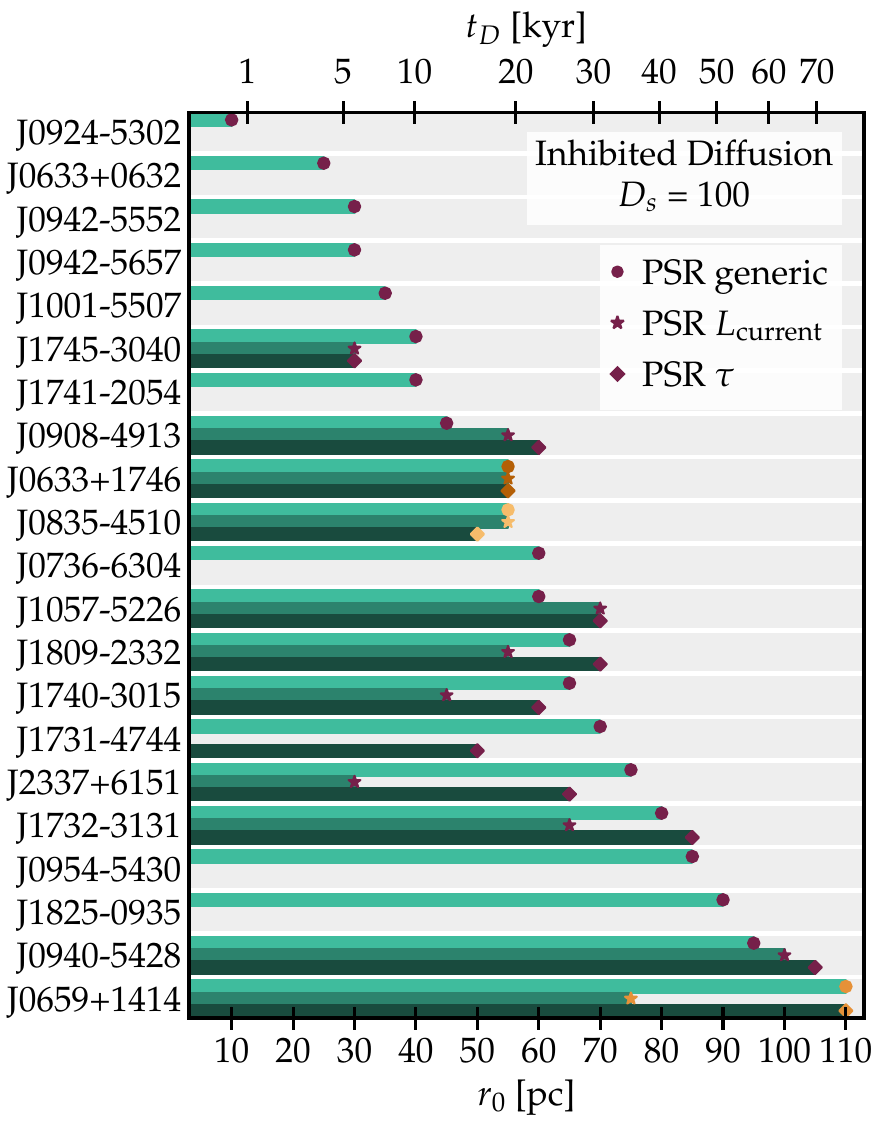}
    \vspace{-0.2cm}
    \caption{The radius of the inner diffusion zone, $r_0$, needed so each ATNF pulsar does not overproduce H.E.S.S. data. Diffusion is suppressed by a fixed value $D_s = 100$. For smaller values of $r_0$ (green shades), the pulsar would overproduce H.E.S.S. data, while gray does not exceed the data. Markers correspond to the critical $r_0$. Vela, Monogem and Geminga are represented by yellow, orange and dark orange markers, respectively. Top axis, $t_D$, shows the time that 10~TeV $e^+e^-$ spend in the inhibited diffusion zone, estimated by ${ r_0 \approx \ell = \sqrt{6Dt} }$. Circles represent the inhibited diffusion zone in our generic model, stars use the current pulsar spindown power $L_\text{current}$ from ATNF, and diamonds use the initial spindown power from the generic model with a spindown timescale $\tau$ that reproduces the current pulsar spindown power.}
    \label{fig: r0 grid}
\end{figure}

Figure~\ref{fig: r0 grid} shows the needed suppression of diffusion for each of the 21~ATNF pulsars that overproduce H.E.S.S. data in one-zone models. In this figure, we use the exact age and distance for each pulsar, but perform our analysis in three scenarios, where we calculate the inhibited diffusion zone using (1) our generic model (light green bars, circle markers), (2) the current ATNF spindown power for each pulsar (medium green bars, star markers), and (3) an initial spindown power of $L_0 = 10^{38}$~erg/s, as in the generic model, but with spindown timescales for each pulsar set such that the spindown power at their characteristic age matches ATNF data (dark green bars, diamond markers). In the latter two cases, we increase the pulsar efficiency from 10\% to 25\% to ensure that the model reproduces TeV observations of Geminga.\footnote{The 10\% efficiency for $e^+e^-$ acceleration in Geminga is calculated from Ref.~\cite{Hooper:2017gtd}, using the Geminga-like model, which predicts Geminga to have a spindown power 2.5$\times$ greater than its observed spindown power. Studies using the current spindown power of Geminga find higher $e^+e^-$ efficiencies~\cite{Hooper:2017gtd, Johannesson:2019jlk}.} App.~\ref{app: model variations} discusses further variations in the pulsar spectra.

We note that size of the inner diffusion region is degenerate with the degree by which diffusion is suppressed. Thus, we fix the suppression factor to $D_s = 100$ and vary $r_0$ until the pulsar no longer overproduces H.E.S.S. data, along with the time that a 10~TeV e$^+$e$^-$ would remain trapped in the region. For comparison, if diffusion is not inhibited, a 10-TeV $e^\pm$ traverses 50~pc in only $\sim 1600$~yr. Fig.~\ref{fig: r0 grid} shows that $e^\pm$ must be trapped for more than 10~kyr in many systems, though the exact timescale varies. For example, 10-TeV $e^\pm$ from Monogem (J0559+1414) must be confined in a low diffusion region for $\gtrsim 70$~kyr, while $e^\pm$ from Geminga (J0633+1746) must only be confined for at least $\sim 20$~kyr.

\emph{We stress that}, throughout this analysis, we compare each pulsar individually to H.E.S.S. data, ignoring any contribution from other pulsars. This makes our result very conservative. In the context of a one-zone model, the total $e^+e^-$ flux from all pulsars in our study would overproduce H.E.S.S. data unless the average pulsar efficiency was less than 0.02\%, which is far smaller than the known $e^\pm$ efficiency of many PWN and TeV halos. We show a figure that compares all pulsar spectra simultaneously to H.E.S.S. in App.~\ref{app: all pulsars}. Even this constraint is conservative, as it omits known contributions from misaligned pulsars~\cite{Linden:2017vvb} and supernova remnants~\cite{Kobayashi:2003kp, Morlino:2021rzv, Sudoh:2023zcu}.

When we use the ATNF spindown power of each pulsar, our conclusions are somewhat weakened by young sources with spindown powers far below our Geminga-centric model. However, we note that there are still 11 pulsars that \emph{individually} overproduce H.E.S.S. data, meaning our result is qualitatively unchanged. Moreover, this model gives our most conservative calculation. When we set $L_0 = 10^{38}$~erg/s for all pulsars, and calculate their individual spindown timescales, our results fall between the Geminga-centric model and ATNF-normalized models, with 12 pulsars that overproduce H.E.S.S. data.

Many studies have investigated the potential for a  misaligned pulsar to be discovered fortuitously close to Earth~\cite{Lloyd:2019rxg, Baryakhtar:2017dbj}. In Ref.~\cite{Linden:2017vvb}, we showed that TeV halos provide a new method for detecting these pulsars. However, the observation of TeV halos around multiple sources did not \emph{require} that all pulsars be surrounded by regions of inhibited diffusion (see Ref.~\cite{Profumo:2018fmz}). Here, we have shown that $e^\pm$ must be trapped for a long duration around every energetic pulsar, which in turn requires that every energetic pulsar powers a bright PWN or TeV halo.

In Appendix~\ref{app: selected ATNF pulsars}, we note that spatially extended emission has already been detected around 13 of the 21 nearby pulsars that overproduce H.E.S.S. data in a one-zone model, including every system in the better-observed Northern hemisphere. Moreover, HAWC has recently shown that TeV halos may be common among middle-aged pulsars~\cite{Albert:2025gwm}. These observations will soon allow us to probe or exclude the existence of undetected nearby middle-aged pulsars.

\section{Summary and Outlook}\label{sec: summary and outlook}
In this \textit{article}, we show that H.E.S.S. $e^\pm$ data demands that a zone of inhibited diffusion must exist around every pulsar that is powerful and close enough to Earth. While observations of both PWN and TeV halos around pulsars clearly show that pulsars are able to accelerate $e^\pm$ to hundreds of TeV, the steep fall of the $e^\pm$ flux above 1~TeV rules out the possibility that these $e^\pm$ travel efficiently to earth, as suggested in one-zone diffusion models. Instead, these observations require a zone of inhibited diffusion around pulsars which allows the $e^\pm$ to lose energy in the vicinity of the pulsar before reaching Earth. Our results further suggest that any pulsar powerful enough to produce high-energy $e^\pm$ must be a bright source, either in radio, x-ray or $\gamma$-rays, which has either already been detected, or soon will be detected by telescopes such as HAWC and LHAASO, or upcoming instruments such as CTA and SWGO.

\section*{Acknowledgements}
Many thanks to Benedikt Schroer for enlightening discussions on the two-zone diffusion model. We also thank Rub\'en L\'opez-Coto and Pierrick Martin for useful comments, and Kun Fang for helpful comments and for pointing out that J1057-5226 was missing in the previous version of the paper. IJ acknowledges support from the Research grant TAsP (Theoretical Astroparticle Physics) funded by INFN, and Research grant ``Addressing systematic uncertainties in searches for dark matter'', Grant No.\ 2022F2843L, CUP D53D23002580006 funded by the Italian Ministry of University and Research (\textsc{mur}). TL is supported by
the Swedish Research Council under contract 2022-04283 and the Swedish National Space Agency under contract 117/19. TL also acknowledges sabbatical support from the Wenner-Gren foundation under contract SSh2024-0037. This project used computing resources from the National Academic Infrastructure for Supercomputing in Sweden (NAISS) under project NAISS 2024/5-666.

% --------------------------------------------------

% --------------------------------------------------

% \newpage
\appendix
\setcounter{equation}{0}
\setcounter{figure}{0}
\setcounter{section}{0}
\setcounter{table}{0}
\makeatletter
\renewcommand{\theequation}{\thesection.\arabic{equation}}
\renewcommand{\thefigure}{\thesection.\arabic{figure}}
\renewcommand{\thetable}{\thesection.\arabic{table}}

\section{Selected ATNF Pulsars}\label{app: selected ATNF pulsars}
In our analysis, we include every pulsar found in the ATNF pulsar catalog~\cite{Manchester:2004bp, ATNF_website} that is within 1.5~kpc of Earth and less than 1~Myr old. In Table~\ref{tab: pulsars}, we list the 21 pulsars that individually overproduce the H.E.S.S. $e^+e^-$ in a one-zone diffusion model (Fig.~\ref{fig: pulsars vs HESS}, top panel), together with observations of an inhibited diffusion zone, which can be a supernova remnant (SNR), pulsar wind nebula (PWN) or TeV halo. For more than half of these sources, a zone of inhibited diffusion has already been discovered. Several of these sources where inhibited diffusion has already been measured (J0633+0632, J0736-6304, J0954-5430 and J1741-2054), do not appear to have inhibited diffusion zones when using our model with the ATNF spindown powers (see Fig.~\ref{fig: r0 grid}), indicating that our analysis is conservative as discussed in the main text. Furthermore, our analysis predicts strong inhibition around J0940-5428, while no such zone has yet been discovered (see Fig.~\ref{fig: r0 grid}).

We note that all pulsars considered here are created in core-collapse supernovae, which means that our results are not directly applicable to millisecond pulsars (MSPs). None of the pulsars in our selected distance and age ranges are classified as MSPs in the ATNF catalogue.

\begin{table*}[tbp]
\centering
\renewcommand{\arraystretch}{1.3}
\begin{tabular}{||c|c|c|c||}
\hline
Pulsar     & Age [kyr] & Distance [pc]  & Inhibited Diffusion Zone   \\
\hline\hline
J0633+0632 & 59.2      & 1355           & PWN~\cite{Danilenko:2020axr}    \\
J0633+1746 & 342       & 190            & TeV halo~\cite{HAWC:2017kbo}    \\
J0659+1414 & 111       & 286            & TeV halo~\cite{HAWC:2017kbo}    \\
J0736-6304 & 507       & 103            & TeV halo candidate~\cite{Lloyd:2019rxg}\footnote{Ref.~\cite{Lloyd:2019rxg} finds hints for an extended $\gamma$-ray emission around J0736-6304 in the context of a different analysis.}    \\
J0835-4510 & 11.3      & 280            & TeV halo~\cite{Sudoh:2019lav}    \\
J0908-4913 & 112       & 1000           & SNR, PWN~\cite{Johnston:2021rkc}    \\
J0924-5302 & 335       & 513            & --    \\
J0940-5428 & 42.2      & 377            & --    \\
J0942-5552 & 464       & 300            & --    \\
J0942-5657 & 323       & 406            & --    \\
J0954-5430 & 171       & 433            & PWN candidate~\cite{Ding:2020wyk}    \\
J1001-5507 & 441       & 300            & --    \\
J1057-5226 & 535       & 93             & PWN~\cite{Posselt:2015bra}, TeV halo~\cite{Wach:2025fqf}      \\
J1731-4744 & 80.4      & 700            & --     \\
J1732-3131 & 111       & 641            & --    \\
J1740-3015 & 20.6      & 400            & PWN, TeV halo~\cite{2010AIPC.1248...25K}    \\
J1741-2054 & 386       & 300            & PWN~\cite{2009ApJ...705....1C, Marelli:2014ioa}    \\
J1745-3040 & 546       & 200            & SNR candidate~\cite{Aharonian:2008gw}    \\
J1809-2332 & 67.6      & 880            & SNR, PWN~\cite{Van_Etten_2012}    \\
J1825-0935 & 233       & 300            & --    \\
J2337+6151 & 40.6      & 700            & SNR~\cite{1993Furst}    \\
\hline
\end{tabular}
\caption{The ages and distances of the 21 pulsars which individually overproduce the H.E.S.S. $e^+e^-$ flux in our one-zone diffusion model for an $e^+e^-$ injection efficiency of 10\%. We list potential sources for the inhibition of diffusion (SNR, PWN, TeV Halos), but note that multiple physical effects could be present that combine to inhibit the efficiency of diffusion~\cite{Mukhopadhyay:2021dyh, Fang:2019iym, Bourguinat:2025kja}.}
\label{tab: pulsars}
\end{table*}

\section{Reduced $e^+e^-$ Efficiency}\label{app: reduced efficiency}
\begin{figure*}[tbp]
\centering
\begin{minipage}[t]{0.47\textwidth}
\includegraphics[width=0.86\textwidth]{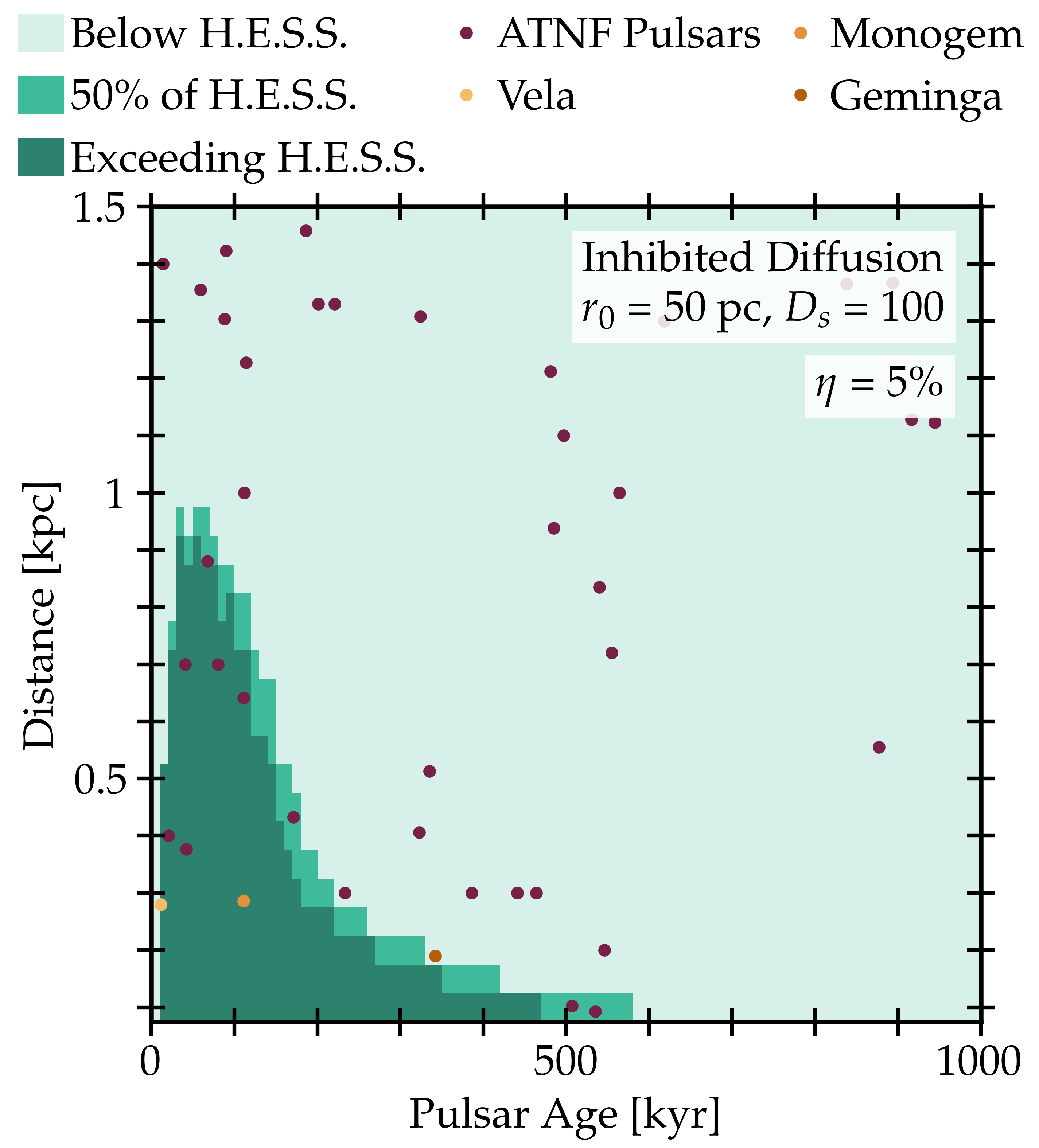} 
\end{minipage}
\hfill
\begin{minipage}[t]{0.47\textwidth}
\includegraphics[width=0.86\textwidth, trim={0 0 0 2.6cm},clip]{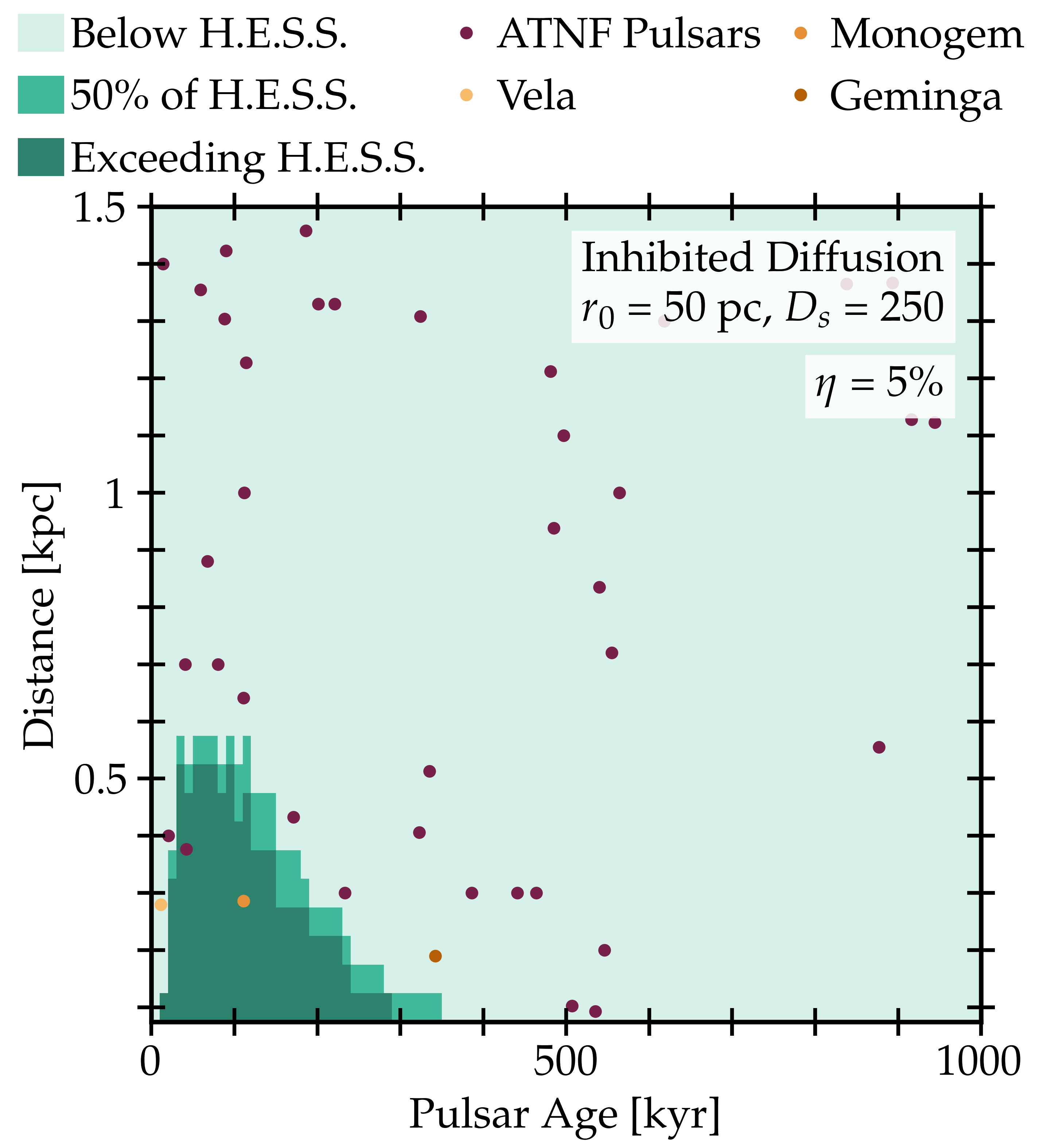} 
\end{minipage}
\vfill
\begin{minipage}[t]{0.47\textwidth}
\includegraphics[width=0.86\textwidth, trim={0 0 0 2.6cm},clip]{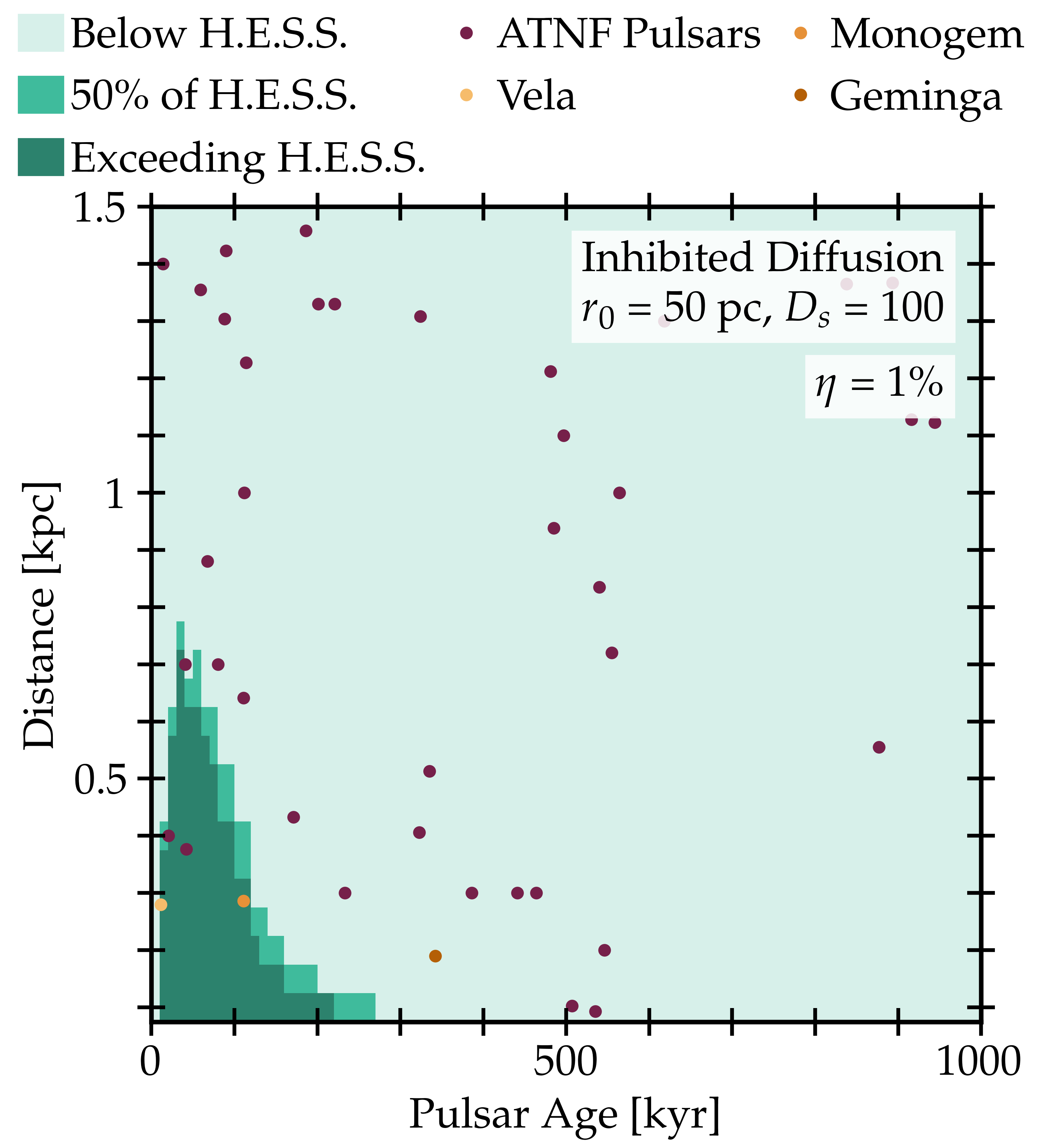} 
\end{minipage}
\hfill
\begin{minipage}[t]{0.47\textwidth}
\includegraphics[width=0.86\textwidth, trim={0 0 0 2.6cm},clip]{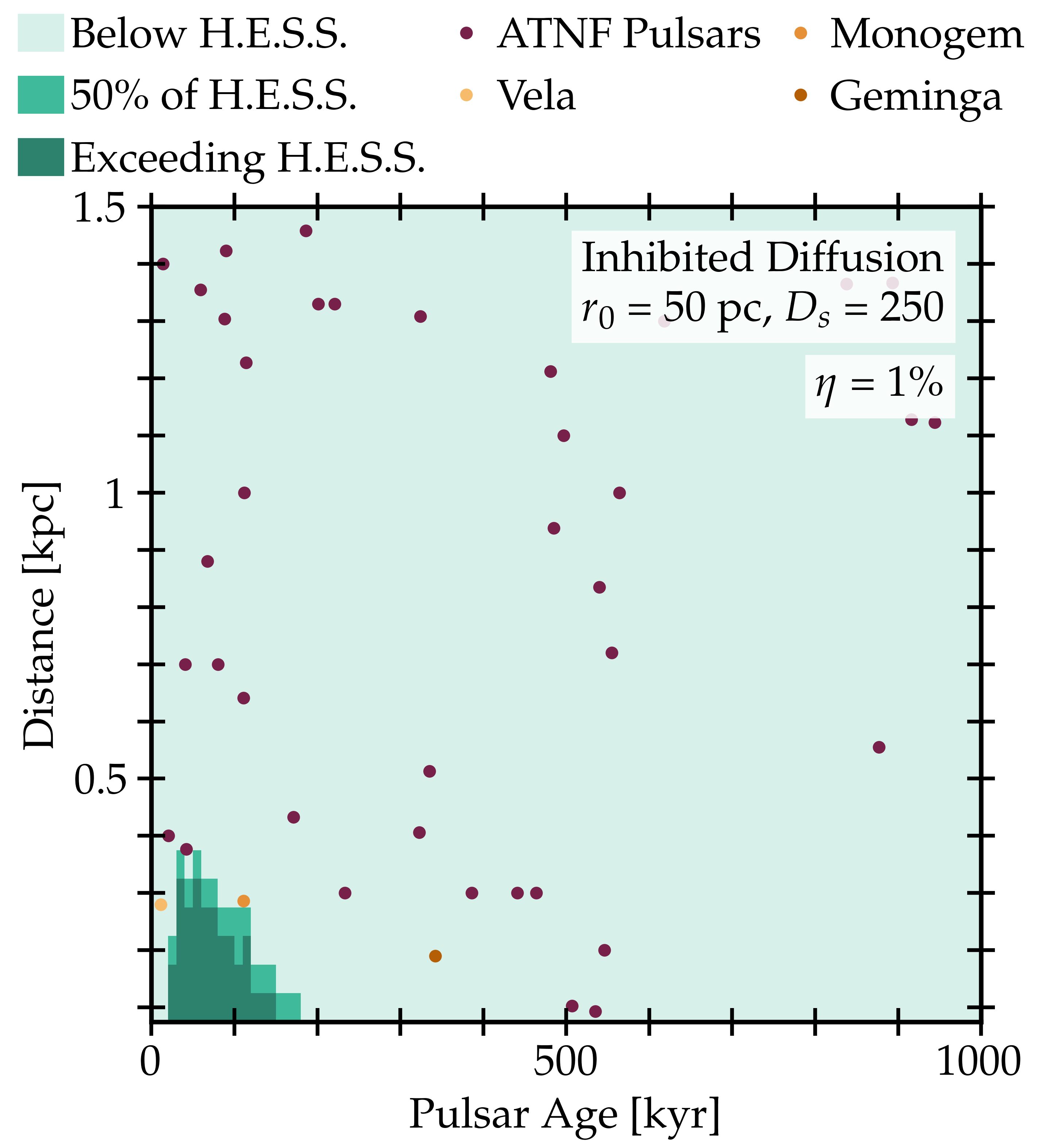} 
\end{minipage}
\caption{Same as Fig.~\ref{fig: pulsars vs HESS} but with a pulsar efficiency of $\eta = 5\%$ (top panels) and $\eta = 1\%$ (bottom panels) for the two-zone diffusion model.}
\label{fig: pulsars vs HESS two-zone 0.01}
\end{figure*}

In our standard scenario in the main text, we assume an efficiency of converting spindown power into $e^+e^-$ pairs of $\eta = 10\%$, based on estimates of Geminga~\cite{Hooper:2017gtd}. However, if Geminga is unusually bright, a lower efficiency might be favoured for other pulsars. We therefore also compare our pulsar models to the H.E.S.S. data assuming efficiencies of 5\% and 1\%. Fig.~\ref{fig: pulsars vs HESS 0.01 eff} shows both scenarios in the one-zone diffusion model. Figure~\ref{fig: pulsars vs HESS two-zone 0.01} shows the pulsar population for the two choices of the two-zone diffusion model, as in Fig.~\ref{fig: pulsars vs HESS} for the 5\% and 1\% efficiencies, respectively. As can be seen, even for a significantly lower pulsar efficiency, that results in a lower flux, the one-zone diffusion model overproduces the data for 14 and 5 sources in the 5\% and 1\% scenarios, respectively, requiring a zone of inhibited diffusion.

\begin{figure*}[tbp]
\centering
\begin{minipage}[t]{0.48\textwidth}
\includegraphics[width=1\textwidth]{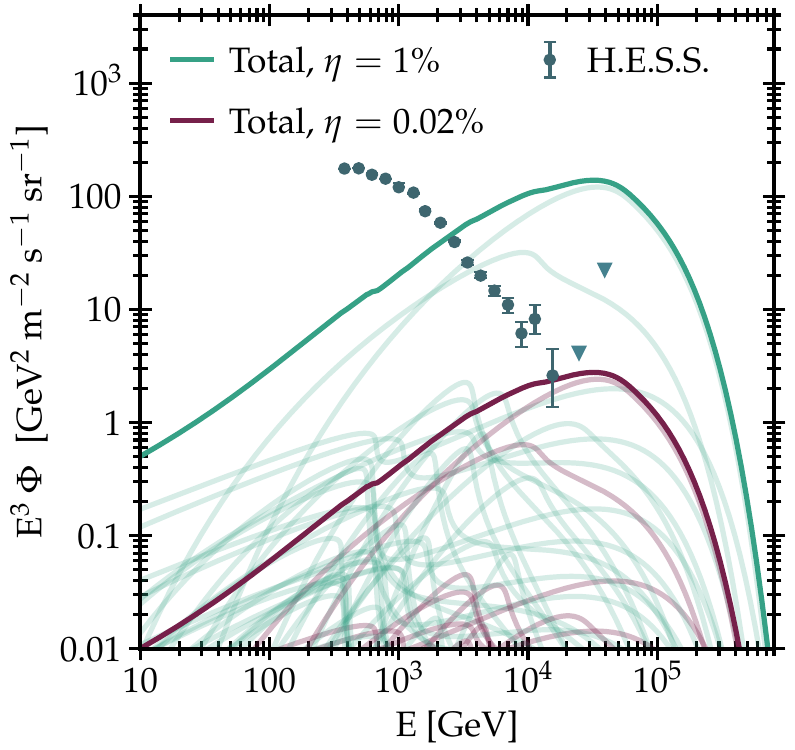} 
\caption{The flux of each of our 41 selected pulsars (light green), and their combined flux (green) compared to the H.E.S.S. data (dark blue). To calculate each pulsar spectrum, we use our model that is based on their current ATNF spindown power, and set the efficiency to a very conservative 1\%. Note that this underproduces the flux of Geminga by a factor of 2.5 (see main text). Even in this conservative scenario, the combined flux strongly overproduces the H.E.S.S. data. For comparison, we show the combined flux with an efficiency lowered further to 0.02\% (purple), which is required to not overproduce the H.E.S.S. data.}
\label{fig: all pulsars vs HESS}
\end{minipage}
\hfill
\begin{minipage}[t]{0.48\textwidth}
\includegraphics[width=0.98\textwidth]{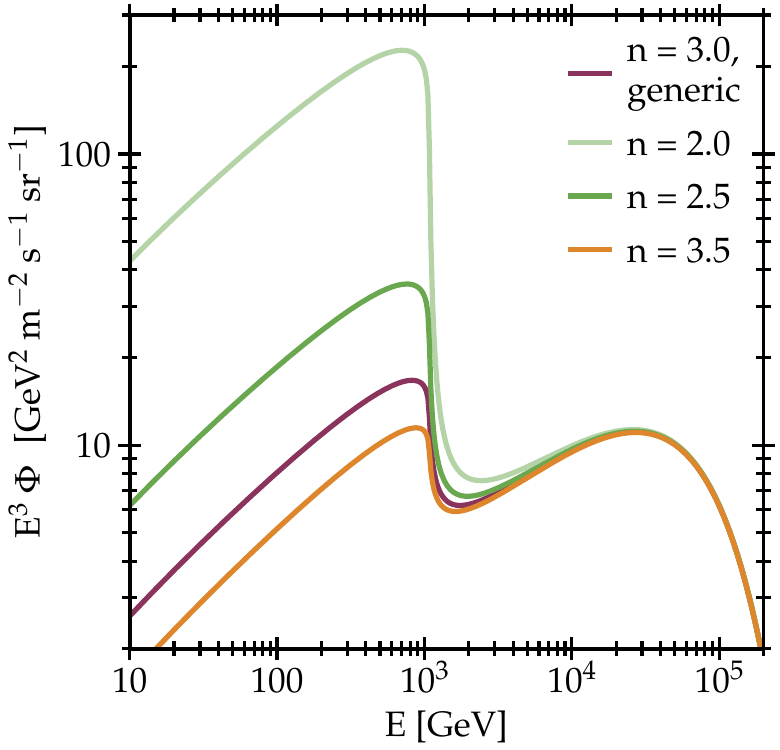} 
\caption{The generic one-zone diffusion model with a braking index of $n=3$ (purple), as assumed throughout this analysis, with alternative assumptions. For smaller braking indices (light and medium green), the low-energy part of the flux is enhanced, as more $e^+e^-$ are injected early on, while for a larger braking index (orange), the slower release of $e^+e^-$ reduced the low-energy flux. Importantly, the high-energy part of the flux is unchanged, as it depends on the most recent injection history, that is fitted to radio observations.}
\label{fig: braking indices}
\end{minipage}
\end{figure*}

\section{Simultaneous Comparison of Pulsars to H.E.S.S.}\label{app: all pulsars}
Throughout our analysis, we only compare each pulsar \textit{individually} to the H.E.S.S. data, \textit{i.e.} we ignore contributions from other pulsars, and do not include any background model of other contributions. This makes our analysis extremely conservative. In Figure~\ref{fig: all pulsars vs HESS}, we show the combined flux of all the 41 pulsars that are within 1.5~kpc from Earth in the one-zone diffusion model (green), to the H.E.S.S. data. We choose the pulsar model based on the current ATNF spindown power for each pulsar, and set the efficiency to a conservative 1\% -- which underproduces Geminga's flux by a factor of 2.5 (and also the fluxes of many other pulsars such as Monogem). Even in this conservative scenario, the combined flux of nearby pulsars significantly overproduces the H.E.S.S. data. For comparison, we show also the combined flux with an efficiency lowered to 0.02\% (purple), which would be required for all pulsars in order to not exceed the H.E.S.S. data. This indicates the importance of the contribution from all nearby and distant sources. Our constraints on the size of the low-diffusion region surrounding nearby sources are likely to be even stronger if additional contributions from the astrophysical background (including both other pulsars and coincident sources such as supernova remnants) are included.

\section{Pulsar Model Variations}\label{app: model variations}
The precise spectrum of a pulsar depends on a variety of parameters, many of which are not well known. In our analysis, the highest part of the pulsar spectrum drives the constraints against the H.E.S.S. data. These high-energy $e^+e^-$ have usually been produced only recently by the pulsar, otherwise they would have experienced more significant energy losses. Therefore, our analysis is driven by the very recent injection history of the pulsar, making our results relatively insensitive to the exact choice of the pulsar evolution history.

In particular, throughout our analysis, we have assumed a braking index of $n=3$, corresponding to a standard dipole (see Eq.~\ref{eq: luminosity}). Observations suggest that the braking index is likely smaller for many pulsars~\cite{Xu:2001bp, Livingstone:2005jj, Lyne:2014qqa, Hamil:2015hqa}. Figure~\ref{fig: braking indices} shows the generic pulsar spectrum (purple) for alternative braking indices. For smaller braking indices (light and medium green), the low-energy part of the flux is enhanced, as the pulsar spins down faster, releasing more $e^+e^-$ early on. However, the high-energy part above about 5~TeV remains unchanged, since these high-energetic $e^+e^-$ have only been injected recently, and are thus mostly independent of the injection history, and since the pulsar spindown power is fixed based on current radio observations. Similarly, in the (unusual scenario) where the braking index is larger (orange), the low-energy part of the flux is reduced, while the high-energy part is unchanged. Since the constraints from our analysis are driven by the high-energy flux, the intensity of the low-energy flux is only mildly relevant, and our conclusions are not highly affected. We note that for very small braking indices, the low-energy data would exceed AMS-02 measurements, providing a new and complementary constraint.

\begin{figure*}[tbp]
\centering
\begin{minipage}[t]{0.48\textwidth}
\includegraphics[width=0.86\textwidth]{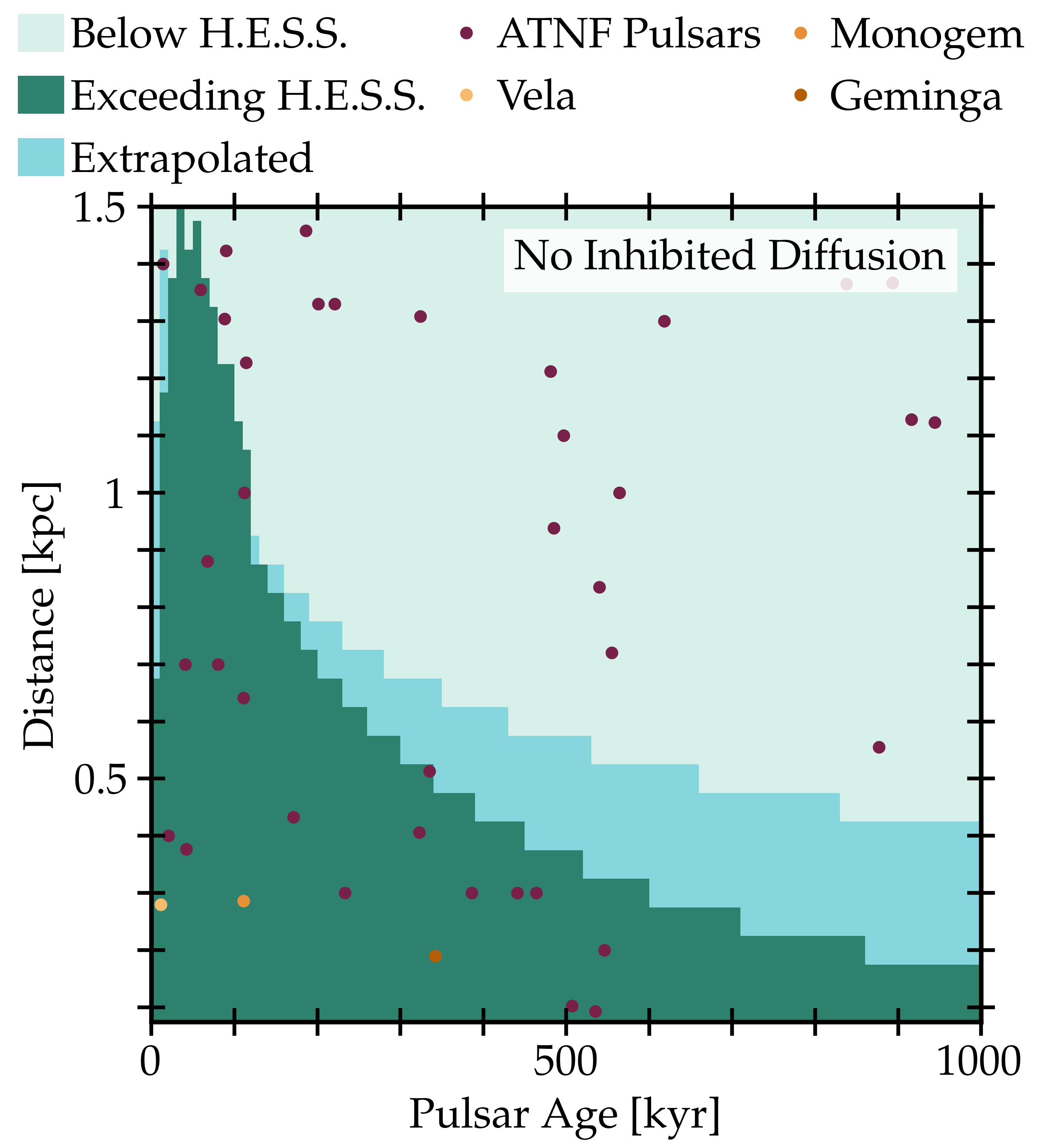} 
\end{minipage}
\hfill\vspace{0.1cm}
\begin{minipage}[t]{0.48\textwidth}
\includegraphics[width=0.86\textwidth, trim={0 0 0 2.6cm},clip]{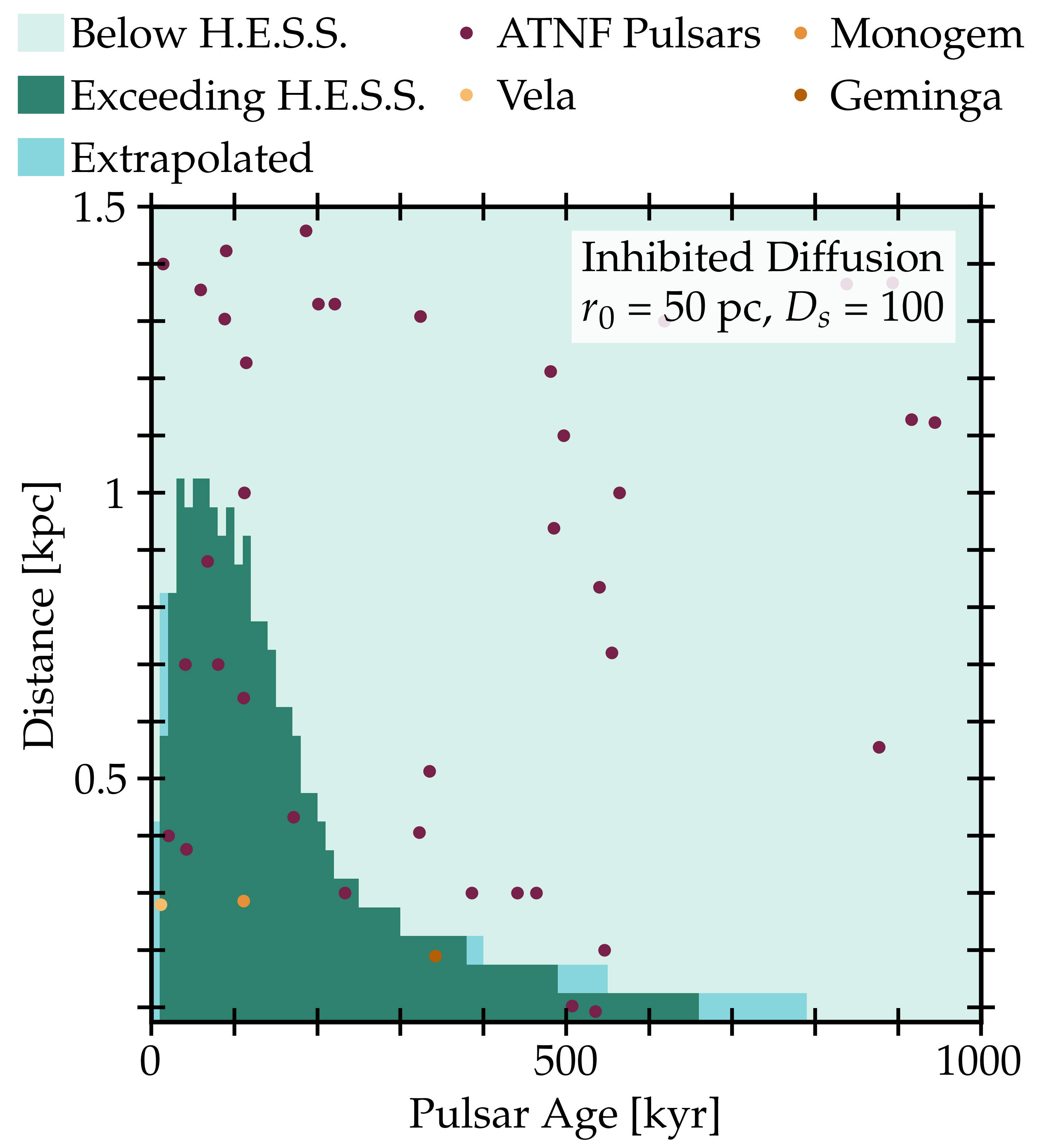} 
\end{minipage}
\hfill\vspace{0.1cm}
\begin{minipage}[t]{0.48\textwidth}
\includegraphics[width=0.86\textwidth, trim={0 0 0 2.6cm},clip]{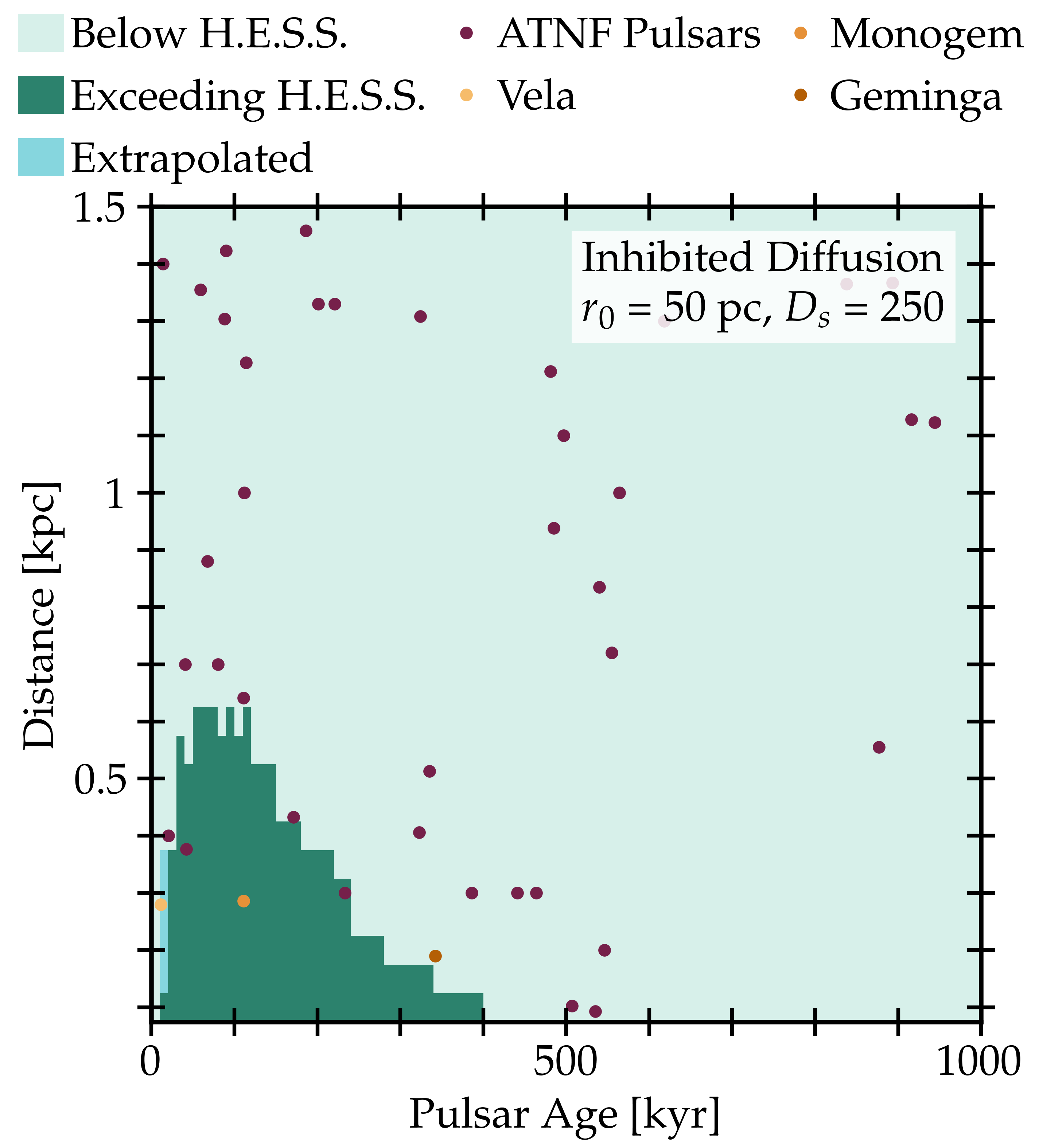}
\end{minipage}
\caption{Same as Fig.~\ref{fig: pulsars vs HESS}, but including a projection of the H.E.S.S. data following a powerlaw with $\Gamma = 4.49$~\cite{HESS:2024etj}, that extends the exclusion of pulsars that would exceed the H.E.S.S. data at energies above the H.E.S.S. range, represented by light blue regions.}
\label{fig: pulsars vs HESS with extrapolation}
\end{figure*}

\section{Projection to Future Telescopes}\label{app: projection to future telescopes}
Upcoming telescopes such as the Cherenkov Telescope Array (CTA)~\cite{CTAwebsite, Knodlseder:2020onx} are expected to measure the $e^+e^-$ flux even further to energies of hundreds of TeV. This will allow us to further constrain the $e^+e^-$ flux from pulsars that dominantly contribute at energies of hundreds of TeV.

We project the potential constraints based on future CTA observations of the high-energy $e^+e^-$ flux by extrapolating the H.E.S.S. data to $\sim$ PeV by following the powerlaw provided by~\cite{HESS:2024etj} that describes the flux for energies above the 1-TeV break with a spectral index of $\Gamma = 4.49$.

We show our results in Figure~\ref{fig: pulsars vs HESS with extrapolation}, which is similar to Fig.~\ref{fig: pulsars vs HESS}, except that the light blue regions represent pulsars that exceed the $e^+e^-$ in the extrapolated energy range. We find that CTA will be able to significantly constrain a much larger population of relatively old pulsars, though, at present there are no known pulsars which inhibit the age- and distance- range probed by CTA observations.

\bibliography{ref}

\end{document}